\documentclass[12pt]{iopart}
\usepackage{cite}
\usepackage[colorlinks, citecolor = blue, urlcolor = blue]{hyperref}
\usepackage[utf8]{inputenc}
\usepackage{graphicx}
\begin{document}
\title{Binding energies of molecular solids from fragment and periodic approaches}

\author{Jaroslav Hofierka$^1$\footnote{Present address:
School of Mathematics \& Physics, Queen's University Belfast, Belfast BT71NN, Northern Ireland, UK}, Jiří Klimeš$^1$}

\address{$^1$Department of Chemical Physics and Optics, Faculty of Mathematics and Physics, 
Charles University, Prague, Czech Republic}

\date{\today}

\begin{abstract}
We calculate binding energies of four molecular solids using the 
Hartree-Fock (HF) and second-order M{\o}ller-Plesset perturbation theory (MP2). 
We obtain the energies within many-body expansion (MBE) as well as using 
periodic boundary conditions (PBC) to compare both approaches.
The systems we study are methane, carbon dioxide, ammonia, and methanol.
We use tight convergence settings to obtain the binding energies with a high precision,
we estimate the uncertainties to be only few tenths of percent.
We discuss several issues that affect the quality of the results and which need to be considered
to reach high precision for both MBE and within PBC.
For example, HF as well as MP2 energies within PBC benefited from the use of real-space
Coulomb cut-off technique,
the convergence of energies within MBE was improved by modifying the order of summation.
Finally, numerical noise made the evaluation of some of the MBE contributions difficult
and the effect was reduced by using smaller basis sets for the less critical terms.
\end{abstract}

\maketitle

\section{Introduction}

Molecular solids are materials important in many areas of science and industry, 
for example as pharmaceuticals. 
However, it is often difficult to obtain their binding energy reliably from theoretical approaches
due to the need to describe accurately both inter- and intra-molecular interactions \cite{price2009}.
For example, the electron density of the molecule and its response to the environment
as well as electron dispersion interactions (van der Waals forces) need to be
accounted for reliably.

The natural way to treat molecular solids is by using periodic boundary conditions (PBC).
Currently, density functional theory (DFT) approximations are widely used within PBC to study 
molecular solids \cite{reilly2016}.
However, due to issues related to, e.g., charge delocalization, the reliability of 
semi-local DFT approximations can be insufficient for properties such as energy differences between polymorphs 
or solid phases even when dispersion corrections are employed \cite{santra2013,reilly2013}.
Hybrid DFT functionals reduce the delocalization errors and lead to more reliable energy differences
\cite{santra2013,reilly2013}.
Despite the higher accuracy, errors in binding energies can still have considerable spread for systems
with similar binding character \cite{klimes2016}.
More consistent results can be obtained from correlated methods, even from simpler ones such as second-order
M{\o}ller-Plesset perturbation theory (MP2) \cite{delben2012jctc} 
or the random-phase approximation (RPA) \cite{li2010rpa,macher2014,klimes2015,klimes2016}.
The use of the reference quality coupled clusters scheme within PBC is currently limited 
by its computational cost \cite{booth2013,mcclain2017,liao2021}.
Finally, quantum Monte Carlo (QMC) was shown to offer a very high accuracy \cite{zen2018}.
However, QMC as well as the other correlated methods usually require
a careful set-up by an experienced user.

As an alternative, the many-body expansion (MBE) can be used to obtain binding energies 
of molecular solids \cite{gillan2013,yang2014bz} as well as other properties \cite{hartman2016}.
Its main advantage is the possibility to use computational methods which are too demanding within PBC,
such as coupled clusters at the reference CCSD(T) level.
Within MBE, the total energy is obtained as a sum of two body energies with non-additive three-, four-, and 
higher-order corrections.
The total computational cost can be affordable if the MBE converges quickly with the size of the fragment
(dimer, trimer, \dots) and with the number of included fragments for each fragment size.
The convergence can be problematic for systems with important long-range electrostatic interactions \cite{richard2014acc}.
This issue can be rectified by, {\it e.g.}, embedding the fragments 
\cite{dahlke2007,bygrave2012,gillan2013} or by using a force-field to account for more distant fragments \cite{wen2011jctc}.
However, it's not clear what precision can be ultimately reached with these approaches \cite{cervinka2018}.
Finally, one can reduce the slow convergence of MBE by starting from a less accurate but 
affordable calculation within PBC and use MBE with a higher-level scheme to improve the result \cite{bludsky2008prb,taylor2012,muller2013}.

In this work we consider the binding energies at the HF and MP2 levels.
For these methods one can find published results that vary by more than 20$\%$ \cite{klimes2015,beran2016}.
A similar situation can be observed for hybrid functionals as they include the HF-like energy.
The origin of such deviations comes likely from the fact that the calculations are computationally
demanding and the total or binding energies can strongly depend on several numerical parameters.
To analyze some of the issues we employ in this work both MBE and PBC to obtain HF 
and MP2 binding energies of four crystals.
For all the systems and methods we try to reach a very high precision of the results.
This allows to analyze the effect of the different parameters and finally reach 
a very good agreement between the data.

\section{Systems}

We selected four systems for our study: methane, carbon dioxide, ammonia, and methanol.
We made this choice to include systems with different binding properties.
Specifically, the long-range electrostatic interactions are negligible for methane, 
but are more important for the other systems.
Moreover, hydrogen bonds can be expected to dominate the binding of methanol.
We expect that the different roles of the various interactions will
affect the convergence behaviour of the calculations.
Finally, the molecules are small, with up to 20 electrons, which simplifies the calculations.

The structures of carbon dioxide and methanol were obtained from the Crystallography Open Database
\cite{grazulis2009,cod}.
For methane, we used a high symmetry structure with fcc packing and the structure of ammonia
is the model $N$ obtained by Boese {\it et al.} \cite{boese1997} based on the structure presented in Ref.~\cite{hewat1979}.
All the structures were used without further optimization and are available in the data repository.
We list their properties in Table~\ref{tab:systems}.

\begin{table}[h]
\caption{The crystals selected for this study, the volumes of the unit cell $V$, the number of molecules 
in the unit cell $Z$, and the dipole moment of the isolated monomer (experimental value from \cite{crc}).\label{tab:systems}}
\begin{indented}
\item[]\begin{tabular}{lcccc}
System         &  Ref.    & $V$ (\AA$^3$) & $Z$    &  $\mu$ (D)\\ 
\hline
Methane        &     & 49.79  & 1 & 0.00 \\
Carbon dioxide &  COD9007643\cite{simon1980}   & 177.88 & 4 & 0.00 \\
Ammonia        & \cite{hewat1979}, \cite{boese1997}   & 135.05 & 4 & 1.47\\ 
Methanol       &  COD5910013\cite{torrie1989}   & 200.53 & 4 & 1.70\\
\end{tabular}
\end{indented}
\end{table}

\section{Methods}

Within the periodic boundary conditions (PBC) approach, the binding energy of a molecular crystal for method M, 
$E_{\rm bind}^{\rm M}$, is given by:
\begin{equation}
E_{\rm bind}^{\rm M}=\frac{E_{\rm cell}^{\rm M}}{Z} -E_{\rm mol}^{\rm M}, \label{eq:pbc}
\end{equation}
where $E_{\rm cell}^{\rm M}$ is the total energy of a unit cell, $Z$ is the number of molecules per unit cell, 
and $E_{\rm mol}^{\rm M}$ is the total energy of an isolated molecule in the gas phase.

The calculations within periodic boundary conditions were performed using the Vienna ab-initio simulation 
package (VASP)\cite{kresse1993, kresse1999}.
The HF and MP2 energies were obtained using the algorithms described in
Refs.~\cite{paier2005,paier2006,marsman2009,grueneis2010}.
Moreover, the so-called direct MP2 energy was evaluated using the random-phase approximation (RPA)
routines in VASP~\cite{harl2008,kaltak2014rpa1,kaltak2014rpa2}.
We used the ``hard" projector-augmented wavefunction (PAW) datasets as supplied with VASP
to minimize the error related to using pseudo-states.
The used basis-set sizes are discussed in the Results section but in general we used orbital cut-offs of at least 600~eV.
The cut-off values for overlap densities and related properties ({\tt ENCUTGW}) were one half of 
the orbital cut-offs.
The real-space cut-off technique \cite{rozzi2006} was used for HF calculations ({\tt HFRCUT}), 
leading to faster convergence with $k$-points and cell volumes, as discussed in Sections~\ref{sec:pbc:hf} and~\ref{sec:pbc:mp2}.
The input files are provided in the repository.
 
The evaluation of the MP2 correlation energy is memory demanding due to the need to store
all the overlap densities between occupied and virtual states for all the $k$-point pairs. 
For large number of $k$-points and basis-set sizes the calculation needs to be run over a large number
of compute nodes so that the total memory is sufficient.
However, this reduces the performance of the code as our systems are rather small and do not scale well.
To bypass this issue, we modified the code to evaluate the MP2 energy only for a single combination
of $k$-points at each run which substantially reduces the memory requirements per calculation.

Within the MBE (or fragment) approach, the binding energy of a molecular crystal is assembled 
from a series of dimer interaction energies of a single arbitrary ``reference" molecule with all 
other molecules in the crystal and higher-order $n$-body ($n=3,4,\dots$) non-additive interaction energies:
\begin{equation}
E_{\rm bind}=\frac{1}{2}\sum_i E^{\rm int}_{0,i} + \frac{1}{3}\sum_{i>j} E^{\rm int}_{0,i,j} + \frac{1}{4}\sum_{i>j>k} E^{\rm int}_{0,i,j,k}+ \dots , \label{eq:mbe}
\end{equation}
where the index $0$ denotes the reference molecule and indices $i,j,k,\dots$ other molecules in the crystal.
The summation thus runs over all non-equivalent dimers, trimers, etc., in a crystal, and
\begin{equation}
\eqalign{
\fl E^{\rm int}_{0,i}=E_{0,i}-E_{0}-E_{i}, \cr
\fl E^{\rm int}_{0,i,j}=E_{0,i,j}-E_{0}-E_{i}-E_{j}-E^{\rm int}_{0,i}-E^{\rm int}_{0,j}-E^{\rm int}_{i,j}, \cr
\fl E^{\rm int}_{0,i,j,k}=E_{0,i,j,k}-E_{0}-E_{i}-E_{j}-E_{k}-E^{\rm int}_{0,i}-E^{\rm int}_{0,j}-E^{\rm int}_{0,k}-E^{\rm int}_{i,j}-E^{\rm int}_{i,k}-E^{\rm int}_{j,k}- \cr
- E^{\rm int}_{0,i,j}-E^{\rm int}_{0,i,k}-E^{\rm int}_{0,j,k}-E^{\rm int}_{i,j,k}.}
\end{equation}
Here, $E_{i}$, $E_{0,i}$, $E_{0,i,j}$, and $E_{0,i,j,k}$ is a monomer, dimer, trimer, and tetramer energy, respectively.

As the summations in Equation~\ref{eq:mbe} are infinite, a cut-off distance $r_{\rm cut}$ needs to be 
introduced above which the contributions are neglected.
We define distance between molecules as the average distance between all the pairs of atoms of the two respective
molecules.
For trimers and tetramers, the total intermolecular distance is a sum of distances of all the pairs of molecules
in the fragment.
The $n$-body energy for a given cut-off distance $r_{\rm cut}$ is then the sum of the all the contributions
with total intermolecular distance below $r_{\rm cut}$.

\begin{figure}[!h]
\begin{center}
 \includegraphics[width=8.5cm]{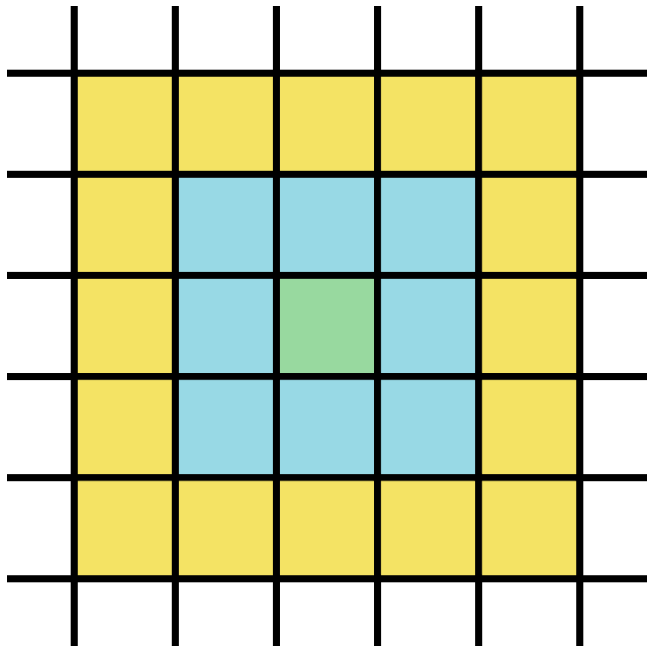}
   \end{center}
   \caption{A sketch of the ``shells" of unit cells used to analyze the contributions of different dimers.}
\label{fig:shell}
\end{figure}

Apart from distance criteria we also define ``shells" of unit cells around the unit cell with the reference molecule.
The unit cell with the reference molecule is ``shell 0", the 26 unit cells around it are ``shell 1",
the next 98 cells belong to ``shell 2" and so on.
This is schematically shown for two dimensions in Figure~\ref{fig:shell} with green, blue, and yellow
used for shell 0, shell 1, and shell 2.
We use then summation based on index of the shell, {\it i.e.}, adding all contributions from molecules in a given
shell, as it leads to improved convergence compared to the standard distance-based criterion.

The set-up of the MBE fragments was done using an in-house library written in Python.
This library also contains functions for extraction and analysis of the results.
Only the energies of symmetry inequivalent fragments were computed.
The symmetry equivalent fragments were obtained by grouping the fragments according to distance and 
comparing the eigenvalues of the Coulomb matrix \cite{rupp2012} for fragments with the same distance, 
similar to the scheme presented by Borca and co-workers \cite{borca2019}.
Note that to sum the energies according to shells we need the full list with all the fragments, 
even symmetry equivalent.
The necessary scripts are in the data repository.

Finite cluster HF and MP2 energies were obtained using the Molpro package \cite{werner2020molpro}.
We used the Dunning's correlation consistent basis sets with augmentation functions \cite{kendall1992}.
The basis-set convergence of interaction energies to the complete basis-set limit was accelerated 
by using the explicitly correlated (F12) variant of 
MP2\cite{kutzelnigg1991A,werner2007,adler2007simple,knizia2009simplified}.
The convergence of HF energies with the basis-set size was improved by the complete
auxiliary basis-set (CABS) correction\cite{adler2007simple,noga2009on}.
The canonical HF and MP2 calculations were run without density fitting.
The MP2-F12 uses density fitting and we have tested both the use of a fitting basis-set
with the same cardinal number as used for orbital basis and a higher cardinal 
number \cite{weigend2002efficient,weigend2005balanced,yousaf2009optimized}.
All energies contributing to an $n$-body term are computed using the basis set of the corresponding $n$-body 
cluster in order to avoid the basis set superposition error (BSSE) \cite{gora2011,rezac2015,modrzejewski2021}.

To increase the precision of the results we set the convergence settings in Molpro as tight
as possible.
Specifically, we used {\tt energy=1.e-16, orbital=1.e-11}, lowered the screening thresholds
for the electron repulsion integrals, and set a tight convergence for F12 {\tt "forb=1.e-15"}.
It was shown that weak convergence criteria can lead to a numerical noise 
that can dominate over the actual energy contributions, especially for three-body or four-body terms\cite{richard2014}.
Indeed, for loose settings we observed that the errors of different fragments do not cancel but 
accumulate so that the binding energy diverges.

\section{Periodic boundary conditions}

\subsection{Hartree-Fock energy}
\label{sec:pbc:hf}

The evaluation of the HF energy in PBC is the simplest one from the four types 
of calculations performed in this work (HF and MP2 in PBC or MBE).
However, it is also a calculation where considerable errors can be introduced.
The reason for this is the Coulomb singularity in the exchange energy which leads to an error proportional 
to the volume of the $\Gamma$-point in the reciprocal space\cite{gygi1986,paier2006}.
The HF energies then converge as $1/N_k$, where $N_k$ is the number of $k$-points or as $1/V$ where $V$ 
is the volume of the simulation cell, $e.g.$, for an isolated molecule.
This error can be essentially mitigated when the Coulomb interaction is cut in the real space outside 
of a specified radius\cite{rozzi2006}.

\begin{figure}[!h]
\begin{center}
 \includegraphics[width=8.5cm]{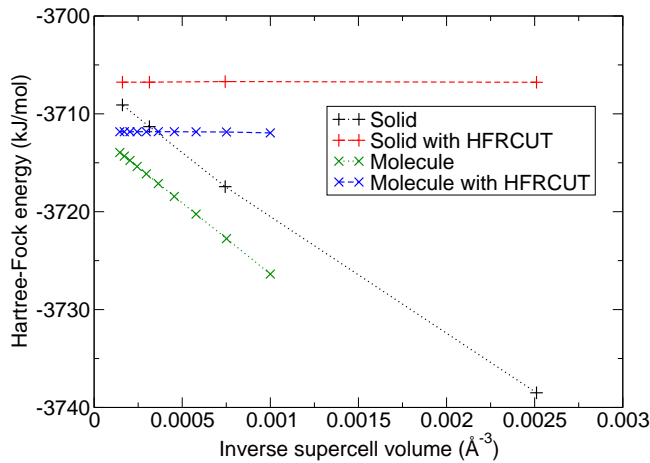}
   \end{center}
   \caption{Convergence of the Hartree-Fock energy for methane solid and isolated molecule as a function 
   of the supercell volume without and with the use of the real-space Coulomb cut-off. 
   For the solid, the supercell volume is obtained as the unit cell volume times the number of $k$-points and
   data between 2$^3$ and 5$^3$ $k$-points are shown. 
   For the molecule, the supercell is the simulation cell.}
\label{fig:pbc:hf1}
\end{figure}

In Figure~\ref{fig:pbc:hf1} we show the HF energies (printed out by VASP) of methane molecule 
and solid for different unit cells and $k$-points, respectively, without and with the use of the real-space cut-off scheme.
As is evident, the cut-off improves the convergence substantially. 
For solid the energy differs from the converged value by around 0.01~kJ/mol already for 2$\times$2$\times$2 $k$-points 
and for molecule a similar convergence is achieved for a cell with 12~\AA\ side.
For the same $k$-point set and cell volume the standard HF energies for solid and molecule have errors 
of 31.7 and 8.4 kJ/mol, respectively. 
Clearly, the standard HF energies need to be extrapolated to infinite volume to remove the error.

We note that this issue also occurs for hybrid DFT functionals and needs to be taken into account 
when energies obtained in different cells are compared, such as for adsorption energies or different phases
of solids.
The problem can be difficult to notice due to the typically smaller amount of Fock exchange in hybrid functionals.
Moreover, if one uses $k$-point set and cell volume for which DFT energies typically converge for insulators 
or semiconductors (supercell side around 20~\AA), the errors partially cancel, even though in an uncontrolled manner.
Also the problem is not present for functionals which screen the long-range part, such as HSE\cite{paier2006}.

We observe a similar improvement of the convergence with $k$-points and volume for the other systems as well when the Coulomb cut-off is used. 
A 2$\times$2$\times$2 $k$-point sampling is sufficient to obtain the energies of ammonia, 
carbon dioxide, and methanol solids to within 0.1~kJ/mol per molecule.
For a 3$\times$3$\times$3 $k$-point grid the energies differ by around 0.001~kJ/mol per molecule from 
a denser $k$-point grid and can be thus considered converged.
Note, however, that this small difference is observed for high basis-set cut-offs (above approx. 1200~eV), 
the energy differences can be larger for smaller basis sets. 

Finally, the energies of molecules that have a non-zero dipole moment need to be extrapolated to infinite cell volume.
This extrapolation corrects the energy of ammonia molecule by only 0.02~kJ/mol compared to a value in a cell with a 21~\AA\ side, 
for methanol the correction is 0.2~kJ/mol for the same settings.
For methane and carbon dioxide we simply used a large enough cell to obtain $E_{\rm mol}^{\rm HF}$. 

The final binding energies were calculated from data obtained with a very high basis-set cut-off, 
2100~eV for methane and methanol and 2000~eV for ammonia and carbon dioxide. 
While the binding energies are converged to within 0.1~kJ/mol for a basis-set cut-off of 1000~eV, 
we opted for a higher cut-off to reduce the uncertainty to around 0.01~kJ/mol.
The final values of the HF energies are summarized in Table~\ref{tab:pbc:hf1}.
As expected, the HF contribution is repulsive for methane and the binding increases in magnitude 
with the increased importance of electrostatic interactions in the systems.

\begin{table}[h]
\caption{The HF binding energies, $E_{\rm bind}^{\rm HF}$, in kJ/mol, the $k$-point grids and cell sizes
used to obtain the energies of solid and molecule, respectively.}
\label{tab:pbc:hf1}
\begin{indented}
\item[]\begin{tabular}{lccc}
System         &  $E_{\rm bind}^{\rm HF}$ (kJ/mol) & $k$-points & cell-size (\AA) \\ 
\hline
Methane        & 5.05  & 5$\times$5$\times$5 & 19 \\  
Carbon dioxide & $-$3.77 & 4$\times$4$\times$4 & 21\\
Ammonia        & $-$11.09& 4$\times$4$\times$4 & 13$-$21 extrap.\\
Methanol       & $-$18.11& 3$\times$3$\times$3 & 15$-$21 extrap.\\
\end{tabular}
\end{indented}
\end{table}

\subsection{MP2 energy}
\label{sec:pbc:mp2}

As with the evaluation of the HF energy, one encounters several possible issues when calculating 
the MP2 energy within periodic-boundary conditions and plane-wave basis-set \cite{marsman2009}.
First, the MP2 energy formula contains a term with zero denominator. 
In VASP, this contribution is evaluated approximately using the derivative of the wavefunction 
with respect to $k$-points ({\tt WAVEDER} file) \cite{gajdos2006}.
Second, additional error occurs if the HF states used in the MP2 calculation were 
not obtained with the Coulomb cut-off technique.
Moreover, the MP2 energy still possesses dependence on $k$-points or cell volume, this dependence
can be reduced by the scheme developed by Liao and Grüneis \cite{liao2016}.
Finally, the MP2 energy depends strongly on the basis set used to expand the orbitals and the overlap 
densities \cite{marsman2009} so that extrapolation to the CBS limit is needed.
We shortly discuss this last point before presenting the results.

The leading term of the MP2 energy errors due to incomplete basis set is proportional to $M^{-1}$, where $M$ is 
the number of basis functions.
For a plane-wave basis set this can be rewritten as $E_{\rm cut}^{-3/2}$ \cite{marsman2009,harl2008,gulans_unp,klimes2014NC}.
This dependence holds for molecule as well as solid and to obtain a converged binding energy one thus needs to extrapolate 
the energies to the complete basis-set limit.
There are, in principle, two ways how to do this.
First, one can obtain the MP2 energies of solid and molecule for different cut-offs, extrapolate each of them to the complete 
basis set (CBS) limit, and subtract them to obtain the binding energy.
Second, one can calculate the binding energies for different cut-offs and extrapolate the binding energy to the CBS limit.
It was shown before that for molecular solids the dependence of the binding energy on the basis-set size can differ from the 
one derived for the total energy and that higher-order errors, 
the first being proportional to $E_{\rm cut}^{-5/2}$, can dominate \cite{klimes2016}. 
We therefore use the second approach where one fits the binding energy directly.

We now turn to the calculation of MP2 correlation energies of isolated molecules.
We evaluate the energies at several basis-set cut-offs to enable extrapolation to the CBS limit, using 600~eV 
as the lowest value and increasing the cut-off in steps of 100~eV up to 1300 or 1400~eV  for methane.
Moreover, for each cut-off, we obtain the energies for different cell sizes starting from 7~{\AA} and increasing 
in steps of 1~{\AA} to enable extrapolation to the infinite cell volume.
Since MP2 memory requirements grow quickly with increasing basis-set or simulation cell size, 
we were only able to use simulation cells up to 12~{\AA} with an orbital cut-off of 1000 or 1100~eV. 
Smaller cells were used for higher cut-offs due to memory limits. 
We find that increasing the orbital cut-off by 100~eV reduces the computationally tractable
cell-size by 1~\AA.

The extrapolation to the infinite cell volume can be done in several ways.
It's important to note that for larger cut-offs (above 1200~eV) we have less datapoints available, 
limiting the accuracy of the fit.
For smaller cut-offs (600 to 900~eV), the data contain numerical noise reducing the accuracy of
the result as well.
After some testing we used to following strategy:
First, we fit the data obtained for the cut-off energy of 1000~eV in cells sized between 8 and 12~\AA\ using a function
$E(x)=E_\infty+Cx^{-6}$\cite{harl2008}.
Second, we fix the coefficient $C$ and fit the data for the other 
orbital cut-offs, again for cells with a side of at least 8~\AA.
This procedure reduces the effect of numerical noise for small cut-offs as all the relevant values are used.

We now discuss the evaluation of MP2 energy for solids which is the most computationally demanding 
part of the calculations within PBC.
The MP2 energy can be divided into two contributions: the direct and exchange MP2, dMP2 and xMP2, respectively.
It is possible to evaluate the first efficiently using the RPA algorithm which scales linearly 
with the number of $k$-points $N_k$ \cite{kaltak2014rpa1,kaltak2014rpa2}.
This reduces the computational time significantly so that the dMP2 term can be evaluated for dense $k$-meshes 
for which the energy is converged to few hundredths of kJ/mol, as discussed below.
Fortunately, the xMP2 energy, for which the compute time scales as $O(N_k^3)$,
converges faster with $N_k$ and similar precision is achieved for coarser meshes.

\begin{figure}[!h]
\begin{center}
 \includegraphics[width=8.5cm]{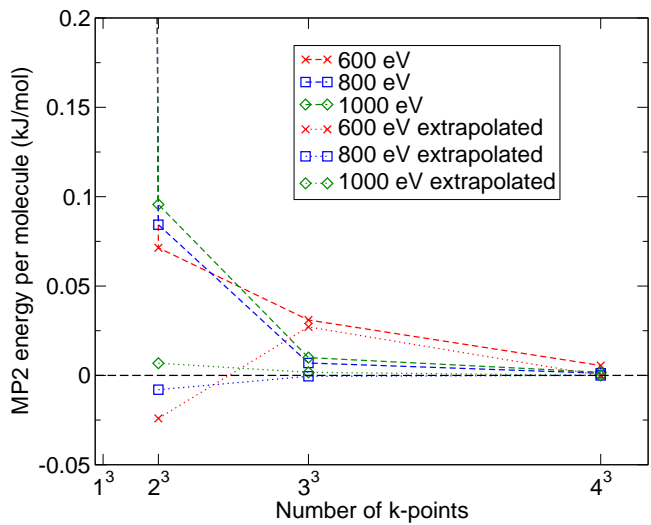}
   \end{center}
   \caption{The direct MP2 correlation energy of ammonia for different $k$-point sets and basis-set cut-offs.
   The values were shifted so that the energies obtained by extrapolation of the $3^3$ and $4^3$ $k$-point data are zero.
   Values are in kJ/mol and per one molecule. The extrapolated values were obtained by a two-point 
   extrapolation assuming that the error decays as $1/N_k^2$ with the number of $k$-points.
   The values for the $\Gamma$-only sampling are close to 5~kJ/mol.}
\label{fig:dmp2:nh3}
\end{figure}

The convergence of the dMP2 energy with the number of $k$-points for ammonia for three different basis 
sets is shown in Figure~\ref{fig:dmp2:nh3}.
The MP2 correlation energy converges as $1/N_k^2$ which can be used to extrapolate the values 
to infinite $k$-point grid, such extrapolations obtained from two adjacent $k$-points are shown with the dotted lines.
All the values in the graph were shifted so that the values obtained by extrapolation of the $N_k=3^3$ and $N_k=4^3$ 
energies are zero.
One can see that already for the 2$\times$2$\times$2 $k$-point set the energy is converged to within 0.1~kJ/mol, 
which might be sufficient for many applications.
Evaluation using the 3$\times$3$\times$3 $k$-point mesh reduces the error by almost an order of magnitude, 
at least for the 800 and 1000~eV cut-offs.
The errors are larger for the 600~eV cut-off which we attribute to numerical noise.
The extrapolation works very well for ammonia, even when only the $\Gamma$-only and 2$\times$2$\times$2
energies are used.
Note that the value for the $\Gamma$-only calculation is at around 5~kJ/mol.
The extrapolation is less efficient for the other systems, we only use it as an additional tool to assess 
the convergence with respect to the $k$-point set.
Also note that all the values are based on HF calculation for which the Coulomb cut-off method was used.
Without it, the errors are 1.4, 0.3, and 0.1~kJ/mol for the 2$\times$2$\times$2, 3$\times$3$\times$3, 
and  4$\times$4$\times$4 $k$-point grids, respectively.

The computational and memory requirements limit again the basis-set cut-off which can be used
with dense $k$-point meshes.
However, the similar dependence of the MP2 error for the 800 and 1000~eV grid, visible in Figure~\ref{fig:dmp2:nh3},
suggests a correction scheme to estimate the energies for higher basis-set cut-offs.
Specifically, to approximate the energy at a high cut-off and dense $k$-mesh $E_{\rm high}^{\rm dense}$, 
we take the energy obtained for the same cut-off and coarse grid $E_{\rm high}^{\rm coarse}$
and correct it with energy difference $\Delta$ between the $k$-points obtained for a lower cut-off
\begin{equation}
\eqalign{E_{\rm high}^{\rm dense}=E_{\rm high}^{\rm coarse}+\Delta\,,\cr
\Delta = E_{\rm low}^{\rm dense}-E_{\rm low}^{\rm coarse}\,.}
\end{equation}
In practice, we obtain the $\Delta$ correction with as high cut-off as possible
and for as many $k$-points as tractable.
The cut-offs and $k$-point sets for which we performed the largest xMP2 and dMP2 calculations 
are summarized in Table~\ref{tab:pbc:delta} together with the value of $\Delta$.
While the value of $\Delta$ depends on the basis-set cut-off, the variations are
usually around or below 0.01~kJ/mol when a cut-off of at least 900~eV is used.

As mentioned previously, the MP2 correlation energy has an error that is proportional to $1/N_{k}^2$.
One can then verify that the error decreases approximately by an order of magnitude when going from 
$n^3$ to $(n+1)^3$ $k$-point grid.
As the values of $\Delta$ are few hundredths of kJ/mol at most, it's highly probable
that going to denser $k$-point meshes would change the energy by less than 0.01~kJ/mol.

\begin{table}[h]
\caption{The largest $k$-point sets, $N_k^{\rm dense}$, used to obtain the xMP2 and dMP2 energies
for each of the systems.
For $N_k^{\rm dense}$ the values were obtained up to a basis-set cut-off value $E_{\rm cut}^{\rm max}$.
For higher cut-offs, the values were taken as the data obtained for $k$-point sampling $N_k^{\rm coarse}$
corrected by the value of $\Delta$ which is the energy difference between the two $k$-point sets
obtained for $E_{\rm cut}^{\rm max}$.}
\label{tab:pbc:delta}
\footnotesize
\begin{tabular}{lcccc@{\hskip12pt}cccc}

System &\multicolumn{4}{c}{xMP2}  & \multicolumn{4}{c}{dMP2}\\
          &  $N_k^{\rm dense}$ & $N_k^{\rm coarse}$ & $E_{\rm cut}^{\rm max}$  & $\Delta$ (kJ/mol) &  $N_k^{\rm dense}$ & $N_k^{\rm coarse}$ & $E_{\rm cut}^{\rm max}$  & $\Delta$ (kJ/mol) \\ 
\hline
Methane        & 4$\times$4$\times$4  & 3$\times$3$\times$3 &  800 & $-0.02$ & 5$\times$5$\times$5  & 4$\times$4$\times$4 &  1000 & $0.02$ \\  
Carbon dioxide & 2$\times$2$\times$2  & 4 & 1100 &  $0.01$ & 4$\times$4$\times$4  & 3$\times$3$\times$3 & 1000  & $0.01$\\  
Ammonia        & 3$\times$3$\times$3  & 2$\times$2$\times$2 & 1000 & $-0.01$ & 4$\times$4$\times$4  & 3$\times$3$\times$3 & 1100  & $-0.01$\\ 
Methanol       & 3$\times$3$\times$2  & 2$\times$2$\times$1 &  700 & $-0.04$ & 3$\times$3$\times$2  & 2$\times$2$\times$1 &  1300 & $-0.002$ \\ 
\hline   
\end{tabular}
\normalsize
\end{table}

The procedure above gives the energies of molecules and solids for a set of basis-set cut-offs and 
we use them to obtain the binding energy, as a function of the cut-off energy.
The binding energy then needs to be extrapolated to the infinite basis-set limit.
As discussed above, the leading order error of the total energies depends as 
$E_{\rm cut}^{-3/2}$ on the plane-wave cut-off, which we observe.
However, the convergence behaviour of the binding energies is different.
Fitting the binding energies with a function $f(x)=E_{\rm \infty} +ax^b$ gives values of exponent $b$ 
between $-2.2$ (carbon dioxide) over $-3$ (methanol and ammonia) to $-4$ (methane).
A fit which assumes $b=1.5$ gives $E_{\rm \infty}$ that differ by around 0.1~kJ/mol for methanol and 
carbon dioxide and by around 0.05~kJ/mol for ammonia and methane from $E_{\rm \infty}$ obtained with varying $b$.
This difference is significant for our purposes.
Moreover, visual inspection of the fit shows that $b=1.5$ is not appropriate, especially 
for methanol where the data have little noise.

The values of exponent $b$ suggest that the next order error, decaying as $E_{\rm cut}^{-5/2}$,
governs the convergence of the binding energy, as for some systems studied in Ref.~\cite{klimes2015}.
Setting $b=2.5$ improves the quality of the fit. 
Moreover, the values of $E_{\rm \infty}$ are within 0.02~kJ/mol of those obtained
with fit where $b$ was a free parameter.
Interestingly, the methane and carbon dioxide data are the most noisy, this is probably 
due to higher symmetry of the crystals compared to methanol or ammonia.

The final values obtained with $b=2.5$ are in Tab.~\ref{tab:pbc:mp1} together with the 
binding energies found for a fixed orbital cut-off of $1300$~eV.
One can see that the extrapolation procedure is necessary for carbon dioxide and methanol 
if one wants to reach a precision of 0.1~kJ/mol or better.

We estimate that the binding energies have overall uncertainties of around 0.1~kJ/mol, 
this value does not include the error of the PAW approximation.
The main sources are extrapolation of energy of isolated molecule to the infinite cell limit,  
errors stemming from the correction scheme for solids and errors of the basis-set extrapolation
for the binding energy.

\begin{table}[h]
\caption{The MP2 binding energies $E_{\rm bind}^{\rm MP2}$ of the studied systems in kJ/mol
extrapolated to the complete basis-set limit and for orbital cut-off equal to 1300~eV.
The fit assumed that the error in the binding energy decreases as $E_{\rm cut}^{-5/2}$
with the plane-wave basis-set cut-off energy $E_{\rm cut}$.}
\label{tab:pbc:mp1}
\begin{indented}
\item[]\begin{tabular}{lcc}
System         &  $E_{\rm bind}^{\infty}$ &$E_{\rm bind}^{1300}$ \\ 
\hline
Methane        & $-14.97$ & $-15.00$\\  
Carbon dioxide & $-25.66$ & $-25.74$\\
Ammonia        & $-23.98$ & $-24.01$ \\
Methanol       & $-37.11$ & $-37.19$\\
\end{tabular}
\end{indented}
\end{table}

\section{Many-body expansion}

\subsection{Two-body contributions}

The two-body energies are the dominant contribution to the MBE of the binding energy.
For Hartree-Fock one observes Pauli repulsion for short distances and electrostatic 
interactions for all distances.
The MP2 correlation falls off as $r^{-6}$ with the distance $r$ between the molecules
which leads to a $r^{-3}$ decay in solid.
To reach high precision of the two-body energy the long-range contributions need to
be accounted for.
To accomplish this, we include all contributions from up to sixth shell.
Moreover, the basis-set incompleteness errors need to be reduced as well, 
which is an issue mainly for MP2.
We start the discussion with results obtained for methane which is the simplest molecule from our set.

Methane has negligible long-range electrostatic contributions and its HF two-body term is thus dominated
by Pauli repulsion. 
In fact, the twelve nearest molecules with intermolecular distance of around 4.2~\AA\ from the reference
molecule contribute around 99\% of the HF two-body energy. 
The basis-set convergence is fast and improved by using the CABS corrections.
For example, while the difference of the HF two-body energies is around 0.04~kJ/mol between the AVTZ and AV5Z
basis sets, it's reduced down to 0.01~kJ/mol when the CABS corrections are added.
We use the HF(CABS) energies obtained with the AV5Z basis set from all the molecules
up to sixth shell to obtain the final two-body energy $E_{\rm HF}^{\rm 2b}=5.265\pm0.002$~kJ/mol.
The uncertainty was estimated based on the difference between the data obtained with different
basis sets and the convergence of the energy on the distance cut-off.

\begin{figure}[!h]
\begin{center}
 \includegraphics[width=8.5cm]{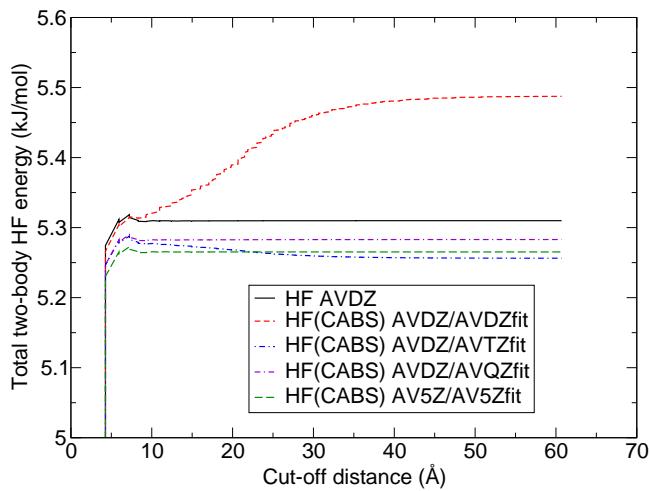}
   \end{center}
   \caption{Convergence of the Hartree-Fock two-body energy of methane without and with the CABS corrections 
   with respect to the cut-off for different basis sets.
   Note that only dimers up to sixth unit cell shell were included, all contributions are included only up
to 23.7~{\AA}.}
\label{fig:2b:me}
\end{figure}

We observed one potential issue when evaluating the HF(CABS) energies of methane.
For the AVDZ basis set, the smallest basis-set used, the CABS corrections critically depend on the fitting basis set. 
The long-range contributions do not go to zero when a double-$\zeta$ fitting basis-sets are used.
This issue is reduced when a larger fitting basis sets are used, as shown in Figure~\ref{fig:2b:me}.

The MP2 two-body energies of methane obtained for all molecules within six shells using different basis sets 
are listed in Table~\ref{tab:2b:me}.
As observed before \cite{bygrave2012}, the errors of MP2-F12 using a basis-set with a cardinal number $N$ are similar to errors
of MP2 using a basis set with a cardinal number $N+1$ or $N+2$.
When the MP2 energies are extrapolated using AV$N$Z and AV$(N-1)$Z basis sets, they have similar errors
to the MP2-F12 in the AV$N$Z basis.
The F12 correction removes the leading order error which decays as $N^{-3}$, the correction 
is less efficient for the higher-order errors.
We have tried if the MP2-F12 energy could be extrapolated as well, assuming that the 
residual error decays as $N^{-5}$.
This seems to work rather well, the extrapolated MP2-F12 value obtained with the AVDZ and AVTZ 
basis sets is within 0.01~kJ/mol from the reference value obtained by extrapolating the AVQZ and AV5Z data.

\begin{table}[h]
\caption{The MP2 energies of methane without and with the F12 corrections for different basis-sets 
and values obtained with a two-point extrapolation assuming the $N^{-3}$, $N^{-5}$ 
error dependence on the cardinal number $N$ for MP2, MP2-F12, respectively. Data are in kJ/mol.}
\label{tab:2b:me}
\begin{indented}
\item[]\begin{tabular}{lcccc}
Basis set        &  MP2  & Extrap. &  MP2-F12 & Extrap.\\ 
\hline
AVDZ & $-13.590$  & $-$ & $ -14.859$ & $-$\\
AVTZ & $-14.783 $ &$-15.286$ &$-15.184 $ & $-15.234 $\\
AVQZ & $-15.058 $ &$-15.259$ &$-15.228 $ & $-15.241 $\\
AV5Z & $-15.149 $ &$-15.245$ &$-15.236 $ & $-15.241 $\\
\end{tabular}
\end{indented}
\end{table}

To estimate the MP2 contributions of dimers in higher shells we fitted the MP2
energy as a function of $r_{\rm cut}$ with  $E_{\infty} +Cr_{\rm cut}^{-3}$.
We used only interval between 12 and 20~{\AA} where the function is sufficiently smooth.
The difference between $E_{\infty}$ and the value obtained by summing contributions
from within the sixth shell is below 0.02~kJ/mol.
Setting a finite distance cut-off $r_{\rm cut}$ for the MP2 interactions, the error would be 
0.1~kJ/mol for $r_{\rm cut}=13$~\AA\ and 0.2~kJ/mol for $r_{\rm cut}=11$~\AA.

We calculate our final value for MP2 two-body energy of methane crystal by taking 
the MP2-F12 energy extrapolated using the AVQZ and AV5Z data and adding the 
correction to infinite $r_{\rm cut}$.
The final value is then $-15.254\pm0.005$~kJ/mol where we the estimated uncertainty
is based on errors of extrapolation to the CBS limit and to infinite $r_{\rm cut}$.

\begin{figure}[!h]
\begin{center}
 \includegraphics[width=8.5cm]{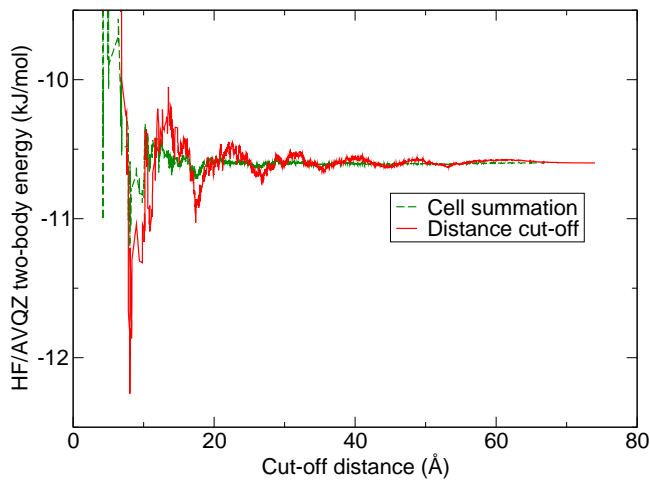}
   \end{center}
   \caption{Convergence of the Hartree-Fock two-body energy for methanol in the AVQZ basis set
   using standard distance based cut-off scheme (Distance cut-off) and its modification where 
   all molecules from a single unit cell are added at the same time (Cell summation). 
   Note that contributions only from molecules
   within sixth shell around the unit cell containing the reference molecule are included, 
   not all contributions above cut-off distance of 30~\AA\ are considered.}
\label{fig:mbe:2b:meoh}
\end{figure}

We now turn to the evaluation of the HF energy for carbon dioxide, ammonia, and methanol.
For these systems the electrostatic interactions are important
and the energies do not converge as quickly with the cut-off distance as for methane.
In fact, the energy as a function of the cut-off distance oscillates as shown for 
methanol in Figure~\ref{fig:mbe:2b:meoh}. 
Note, that we use only the dimers up to sixth shell which means that dimers from larger shells 
are not included in the graph, first such dimer occurs at a distance of around 30~\AA.
One way to deal with the convergence is to parametrize a force-field and use it to include the contributions
above the cut-off distance \cite{beran2010jpcl}.
However, the structure of methanol suggests that the oscillations can be caused by the fact that for 
some cut-off distance the reference molecule interacts more with molecules with an aligned dipole while 
for a different cut-off there are more interactions with molecules with a dipole oriented in the opposite way.
To test this we reordered the summation so that all the molecules from the same unit
cell contribute at the same cut-off distance.
The cut-off is given by the smallest distance to the reference molecule for these molecules.
The modified summation clearly improves the convergence of the HF two-body energy, as illustrated by the 
green line in Figure~\ref{fig:mbe:2b:meoh}. 

While the summation over unit cells improves the convergence, 
for numerical analysis it's beneficial to simplify the summation further.
To this end, we separate the dimers according to the shell of the second molecule and 
sum all contributions within each shell.
The data show a fast convergence with the index of the shell, see Table~\ref{tab:2b:co2prec} 
for carbon dioxide as an example.
For methanol and ammonia, taking 499 two-body contributions from up to second shell 
leads to an error of around 0.1~kJ/mol.
In the case of methanol the standard distance based summation still 
shows errors three times larger for the same number of dimers.

The long-range contributions are needed to reach high precision but their large number leads to 
increased computational effort when large basis sets are used.
One would therefore hope that for large distances a small basis set (such as AVDZ or AVTZ) would be sufficient
to obtain the shell contributions with an error well below 0.01~kJ/mol so that the use of a large basis set could be avoided.
However, it turns out that the contributions of more distant shells have actually larger errors when 
larger basis sets (AVQZ or AV5Z) are used than with the small basis sets.
This is illustrated by the shell contributions obtained for carbon dioxide in Table~\ref{tab:2b:co2prec}.
One can see that for the third and higher shells the contributions are well below 0.01~kJ/mol in the AVDZ
and AVTZ basis sets.
In the AVQZ and AV5Z bases the contributions are positive and sum to almost 0.05~kJ/mol for the AV5Z basis set.
Clearly adding more dimers would further reduce the precision of the result.
This behaviour is a consequence of the numerical errors: to reach a precision of 0.001~kJ/mol for the total 
contribution of the sixth shell, each of the term needs to be evaluated with a precision of around $10^{-7}$~kJ/mol 
or $10^{-10}$~Ha.
Given that the total energies are in the order of $10^6$~kJ/mol or $10^{3}$~Ha, this is approaching 
the limit of double precision arithmetic.
Moreover, reaching the high precision becomes more difficult with the increasing size of matrices that 
need to be transformed or diagonalized during the HF calculation.
That is, the issue is more severe for larger basis sets and also for larger molecules.

\begin{table}[h]
\caption{The Hartree-Fock contributions of different shells in kJ/mol for carbon dioxide along with the number
of dimers in each shell.}
\label{tab:2b:co2prec}
\begin{indented}
\item[]\begin{tabular}{lccccc}
Shell  index       &  \#dimers & AVDZ & AVTZ & AVQZ & AV5Z \\ 
\hline
 0 &   3 &    $-$0.4084  &  $-$0.3642  &   $-$0.3861 & $-$0.4080 \\
 1 & 104 &    $-$3.0995  &  $-$2.8825  &   $-$2.9313 & $-$2.9993 \\
 2 & 392 &    $-$0.0251  &  $-$0.0242  &   $-$0.0225 & $-$0.0226 \\
 3 & 872 &    0.0002  &  $-$0.0001  &    0.0027 &  0.0019 \\
 4 & 1544&    0.0001  &  $-$0.0005  &    0.0033 &  0.0067 \\
 5 & 2408&    0.0001  &  $-$0.0009  &    0.0047 &  0.0171 \\
 6 & 3464&    0.0003  &  $-$0.0011  &    0.0058 &  0.0225 \\
 Sum & 8787  &   $-$3.5321 & $-$3.2736&   $-$3.3234 & $-$3.3817 \\
\end{tabular}
\end{indented}
\end{table}

Given the numerical issues of the large basis sets we have used them only for the shells 
with small indices when obtaining the final results.
The contributions of larger shells were obtained with the AVTZ basis set.
Our final HF two-body energies are summarized in Table~\ref{tab:mbe:all}
and show expected behaviour, the interaction is repulsive for methane and attractive
for the other systems.

\begin{table}[h]
\caption{The HF energies and estimated uncertainties due to finite cut-off and incomplete basis set in kJ/mol for different orders of MBE.}
\label{tab:mbe:all}
\begin{indented}
\item[]\begin{tabular}{lcccc}
System         &  Two-body & Three-body & Four-body & Total \\ 
\hline
Methane        & $5.265\pm0.002$ & $-0.199\pm0.003$ & $0.003\pm0.005$ & $5.07\pm0.01$ \\  %
Carbon dioxide & $-3.43\pm0.01$  & $-0.49\pm0.03$   &   $ 0.14\pm0.03$ & $-3.78\pm0.07$ \\ 
Ammonia        & $-10.75\pm0.01$ & $-0.75\pm0.10$   &  $0.49\pm0.05$  & $-11.01\pm0.16$ \\ %
Methanol       & $-10.60\pm0.01$ & $-7.83\pm0.30$    & $0.30\pm0.10$ & $-18.13\pm0.41$  \\
\end{tabular}
\end{indented}
\end{table}

We now discuss the MP2 contributions to the two-body MBE energies of carbon
dioxide, ammonia, and methanol.
Contrary to the methane crystal, the MP2 two-body energy 
does not converge smoothly enough with the distance cut-off to enable extrapolation to the infinite cut-off. 
The situation is improved by reordering the summation, as with the HF energy, 
but we have decided to base the analysis again on the shell contributions.

As expected, the largest two-body contributions come from the zeroth and first shell, that is from  
molecules either in contact with the reference molecule or in a close distance.
For all three crystals the molecules in the second shell, in which the molecules are at least 8~\AA\ away 
from the reference molecule, still have a significant contribution to the binding:
The MP2 two-body energies are  $-0.35$, $-0.46$, and $-0.58$~kJ/mol for ammonia, carbon dioxide, and methanol, respectively.
For higher shells, the contributions have a similar magnitude for the three systems and sum to approximately $-0.1$~kJ/mol.

We now turn to the basis-set dependence of the MP2 energies as a function of distance.
For simplicity, we compare the energies obtained for different shells.
As with HF, we want to see whether smaller basis-sets are sufficient to reproduce the values
obtained with large basis sets and for which shells this happens.

We use ammonia as an example and show the errors of AVDZ and AVTZ basis sets for different shells
in Table~\ref{tab:mbe:bse}.
The reference data are the MP2-F12/AV5Z two-body contributions listed in the last column.
One can see that using MP2 with AVDZ basis set for the proximate dimers (in zeroth and first shells) leads
to an error of around 5 kJ/mol, close to 20\% of the total two-body energy.
In contrast, using AVDZ for dimers in the second and higher shells leads to a small error of around 
0.02~kJ/mol.
The errors in all shells decrease approximately by a factor of three when the AVTZ basis set is used 
or when the F12 corrections are included in the AVDZ calculation.
For MP2-F12 in the AVTZ basis set the total error is close to 0.13~kJ/mol and predominantly comes
from the first two shells.

The errors in the MP2 two-body terms can be reduced by extrapolation.
In the case of canonical MP2, the error is reduced to around 0.2~kJ/mol when the AVDZ and AVTZ 
values are used.
The result of extrapolation using the AVDZ and AVTZ data is not completely satisfactory, 
in agreement with observations in the literature \cite{takatani2010}.
However, the error becomes essentially marginal, close to 0.01~kJ/mol,
when the MP2-F12 energies in the same bases are extrapolated to the CBS limit.
We again assumed the $N^{-5}$ convergence with the cardinal number of the basis set.
We observe a good performance of the MP2-F12 extrapolation also for methanol, 
where the errors for zeroth and first shells are around 0.01~kJ/mol.
In contrast, when the canonical MP2 energies are extrapolated (still using AVDZ and AVTZ basis sets),
the contributions of the first two shells have each an error of over 0.3~kJ/mol.

\begin{table}[h]
\caption{Basis-set errors of MP2 ($\Delta$ MP2) and MP2-F12 ($\Delta$MP2-F12) 
for two-body contributions from different shells of ammonia crystal with respect to
the MP2-F12/AV5Z reference two-body energies (last column). The extrapolated data use the AVDZ and AVTZ values
and assume $N^{-3}$ and $N^{-5}$ decay of the error with the cardinal number of the basis set $N$
for the canonical and F12-corrected MP2, respectively.
The MP2-F12/AVDZ calculations were performed with quadruple-$\zeta$ density-fitting basis sets.
All energies are in kJ/mol.\label{tab:mbe:bse}}
\footnotesize
\begin{tabular}{cccccccc@{\hskip10pt}c}
       &    &  $\Delta$MP2 & &  & $\Delta$MP2-F12 & & & MP2-F12\\ 
Shell  index&  \#dimers &  AVDZ & AVTZ & Extrap.  & AVDZ  & AVTZ & Extrap. & AV5Z \\
\hline
 0 &   3 & 0.489 & 0.117 &$-0.040$& 0.154 & 0.019& $-0.001$ &$-3.108$  \\
 1 & 104 & 4.328 & 1.419 & 0.193 &  0.760 & 0.107& $0.008$ &$-20.888$ \\
 2 & 392 & 0.013 & 0.004 & 0.001 &  0.011 & 0.002& $0.001$  &$-0.353$  \\
 3 & 872 & 0.002 & 0.001 & 0.000 &  0.001 & 0.000& $0.000$  &$-0.052$   \\
 4 & 1544& 0.001 & 0.000 & 0.000 &  0.001 & 0.001& $0.001$  &$-0.015$  \\
 5 & 2408& 0.000 & 0.000 & 0.000 &  0.001 & 0.001& $0.001$  &$-0.006$   \\
\end{tabular}
\normalsize
\end{table}

Finally, the numerical errors in the HF two-body contributions of higher shells
also affect the subsequent MP2 energies. 
For example, the sixth shell contribution of carbon dioxide becomes positive in the AV5Z basis set. 
The error is, however, below 0.01~kJ/mol and thus barely noticeable.
In any case, we use the large basis set results only from shells with small indices and use 
AVQZ or AVTZ results for the distant dimers.
Our final values are summarized in Table~\ref{tab:mbe:mp2}, the uncertainties come from estimates 
of basis-set incompleteness errors and estimated contributions beyond the sixth shell (below $0.01$~kJ/mol). 

\begin{table}[h]
\caption{MP2 correlation energies and their estimated uncertainties due to finite cut-off and incomplete basis set for different orders of MBE.
All data are in kJ/mol.
}
\label{tab:mbe:mp2}
\begin{indented}
\item[]\begin{tabular}{lcccc}
System         &  Two-body & Three-body & Four-body & Total \\ 
\hline
Methane        & $-15.254\pm0.005$ & $0.255\pm0.002$&  $-0.009\pm0.009$ & $-15.01\pm0.02$ \\  
Carbon dioxide & $-26.64\pm0.02$   & $0.88\pm0.02$  &  $-0.02\pm0.02$ & $-25.78\pm0.06$ \\ 
Ammonia        & $-24.42\pm0.02$   & $0.41\pm0.02$  &  $0.03\pm 0.03$& $-23.98\pm 0.07$\\ 
Methanol       & $-37.54\pm0.02$   & $0.48\pm0.02$  &  $-0.09\pm0.09$& $-37.15\pm 0.13$ \\  
\end{tabular}
\end{indented}
\end{table}

\subsection{Three-body contributions}

The three-body contributions can be sizable and the number of trimers required
to reach convergence can be also high \cite{kennedy2014}.
Moreover, the accurate summation can be problematic due to the numerical issues \cite{richard2014}.
These issues make the three-body contributions critical for the evaluation of the binding energy within MBE.
As with dimers methane shows a different convergence than the other systems and we discuss it first.

It was previously shown that the three-body terms are less sensitive to the choice of the basis set compared with the two-body energies \cite{gora2011,modrzejewski2021}.
For methane, we observe a difference of only 0.002~kJ/mol between CABS corrected HF energies obtained with the AVDZ and AVQZ basis sets.
The difference is approximately 0.005~kJ/mol for standard HF.
The MP2-F12 values also differ by around 0.002~kJ/mol between the AVDZ and AVQZ basis sets and the AVTZ 
is less than 0.001~kJ/mol away from the AVQZ energies.
We use the AVTZ basis set also for the other systems.
One could expect that the stronger electrostatic interactions in the other systems would lead to larger 
basis-set errors but we generally observe only small changes of the binding energy due to the F12 and CABS corrections. 

The convergence of the HF three-body energy with the cut-off distance is fast only for methane.
Here a distance cut-off of 18~\AA\ is sufficient to essentially converge the three-body energy to a value of $-0.199$~kJ/mol.
Adding contributions above this cut-off only varies the energy by around $\pm0.003$~kJ/mol.
The MP2 energy converges as $r_{\rm cut}^{-4}$, in agreement with the expected behaviour.
However, there is little need to perform extrapolation.
The MP2 energy obtained for a cut-off distance 20~\AA\ is within 0.01~kJ/mol of the extrapolated value.
We note that the F12 corrections show some noise which depends on the basis-set size.
However, the noise is only few thousandths of kJ/mol for the small AVDZ basis set 
(with triple-$\zeta$ fitting basis sets in F12).

The stronger electrostatic interactions present in the other systems lead to less systematic convergence, 
similar to the situation observed for dimers.
And as with dimers, the convergence can be improved when the summation is reordered.
When a trimer contribution $E^{\rm int}_{0,i,j}$ is added for some $r_{\rm cut}$, 
we also include contributions of trimers $E^{\rm int}_{0,i,k}$, where the molecules $k$
share the unit cell with the molecule $j$.
In standard summation, the $E^{\rm int}_{0,i,k}$ contributions would be included at higher cut-off distances.
We observe improved convergence not only for methanol and ammonia, but also for carbon dioxide
and both for HF and MP2.

\begin{figure}[!h]
\begin{center}
 \includegraphics[width=8.5cm]{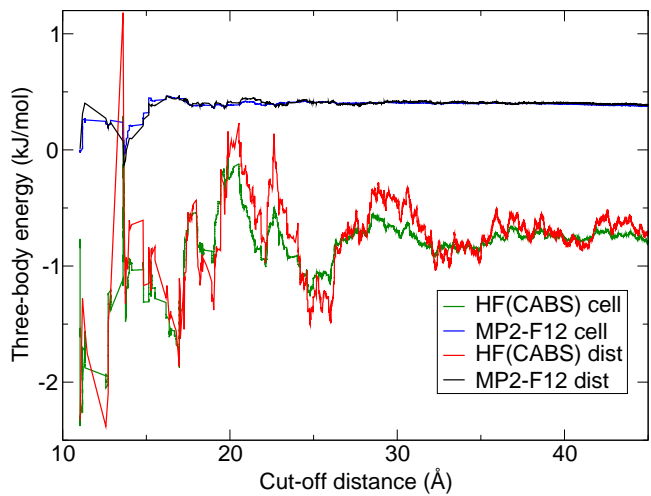}
   \end{center}
   \caption{HF(CABS) and MP2-F12 three-body energies of ammonia as a function of the total intermolecular cut-off distance using the standard 
   cut-off (``dist" data, red and black line) and using the summation where all possible trimers sharing the first two molecules are added 
   at the same time (``cell" data, green and blue line).}
\label{fig:mbe:3b:nh3}
\end{figure}

We demonstrate the improved convergence of three-body HF(CABS) and MP2-F12 energies for ammonia in Figure~\ref{fig:mbe:3b:nh3}.
The standard cut-off based summation is shown with the ``dist" data (red and black line),
the modified summation with the ``cell" values (green and blue line).
For the HF energies, the standard cut-off summation has oscillations of few tenths of kJ/mol even above a cut-off distance of 40~\AA.
The cell summation reduces the range of the values to approximately one half.
Similar improvement can be observed for the MP2-F12 energies, but here the spread of values is in the order 
of hundredths of kJ/mol already for the standard summation.

We were able to obtain the three-body energies in the AVTZ basis set up to a cut-off distance of 40~{\AA} for methanol, 
46~{\AA} for ammonia, and 45~{\AA} for carbon dioxide using the modified summation.
Note that the energies contain also contributions from some trimers with higher total intermolecular distance
due to the use of the modified summation.
We estimate the three-body energy by its average value over the last 5~\AA.
We estimate the uncertainty as one half of the difference of the largest and smallest value of the 
three-body energy on the last 10~\AA.

We list the HF three-body energies and their uncertainties in Table~\ref{tab:mbe:all}.
Even with the modified summation, the data for ammonia and methanol are not tightly converged
and have uncertainties of $0.10$ and $0.3$~kJ/mol.
The convergence is much faster for carbon dioxide, which has no dipole.
The three-body HF energy converges at a cut-off of around 25~{\AA} and increasing
the cut-off further mainly leads to a reduced uncertainty.
We note that the stated uncertainties might be rather conservative. 
We have fitted a simple forcefield that accounts only for polarization to the methanol 
three-body energies and found that increasing $r_{\rm cut}$ to 70~{\AA}, while using the cell summation, 
changes the three-body energy by only $+0.04$~kJ/mol.

The MP2 energies show a fast convergence, averages taken over intervals of 5~{\AA} are essentially converged 
to within few hundredths of kJ/mol when the lower limit is at least 15~{\AA}.
However, to converge the uncertainty or spread of the values to a similar range requires a cut-off
of around 30~{\AA}.
Importantly, the MP2 data of carbon dioxide and ammonia start to show effect of numerical noise 
above a cut-off of 30~{\AA}.
We therefore use the average from interval between 25 and 30~{\AA} as the final value.
The values of methanol do not exhibit such behaviour and we use all the values up to $r_{\rm cut}=40$~\AA.

Overall, we find that all the HF contributions are attractive and all MP2 three-body energies are repulsive.
All the three-body energies are below 1~kJ/mol, apart from the HF value for methanol which is almost as large as the two-body term.
This is due to strengthening of hydrogen bonds in trimers.

\subsection{Four-body contributions}

The four-body terms are smaller compared to the three-body contributions 
yet they can't be completely neglected \cite{gora2011,heindel2020}, 
at least for Hartree-Fock. 
Their magnitude and importance is expected to be smaller for MP2 compared to HF. 
However, the precise evaluation of four-body terms is difficult due to numerical issues.
We observed for the two-body terms that the numerical issues grow with the basis-set size. 
We therefore used the AVDZ basis-set together with quadruple-$\zeta$ fitting basis sets 
for CABS and F12 to evaluate the four-body terms.
Fortunately, small basis sets do not introduce significant errors in the four-body contributions\cite{gora2011}.

We again evaluate the four-body terms using the cell summation.
Here all tetramers sharing the first three monomers and with the last monomer in the same unit cell
are added to the energy at the same $r_{\rm cut}$.
The cut-off distance at which is the contribution added is given by the shortest total intermolecular
distance of all such fragments.

We were able to evaluate the four-body contributions using a cut-off distance of 40~{\AA} 
for methane, ammonia, and methanol and 45~{\AA} for carbon dioxide.
These cut-offs required around 10 000 individual four-body terms.
For the estimates of HF energies we take the average of HF energies on the last 5~\AA, 
the uncertainty is again given by the spread of the values.

The final HF values and their uncertainties due to cut-off are summarized in Table~\ref{tab:mbe:all}. 
All the contributions are destabilizing and reach up to 0.49~kJ/mol for ammonia.
Interestingly, while the four-body contributions are smaller than the three-body energies, 
the uncertainties do not decrease proportionally, apart from ammonia.
While increasing the cut-off further could reduce the uncertainties, the values are close
to the limits given by the numerical precision.
For example, we see that increasing the cut-off for methane deteriorates the precision of the result
and we use only data up to a distance of 30~{\AA}.

\begin{figure}[!h]
\begin{center}
 \includegraphics[width=8.5cm]{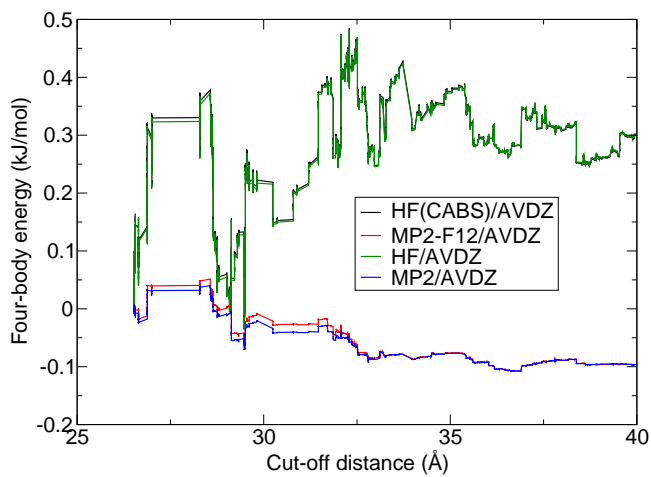}
   \end{center}
   \caption{Four-body energies of methanol for different methods evaluated with the AVDZ basis set.
     The cell summation was used to evaluate the energy dependence on the cut-off distance.}
\label{fig:mbe:4b:meoh}
\end{figure}

Figure~\ref{fig:mbe:4b:meoh} shows the convergence of four-body terms for methanol
and illustrates some of the mentioned points.
For example, there is only a small difference between HF(CABS) and HF values,
as well as between MP2 and MP2-F12.
This shows that the AVDZ basis-set is sufficient to evaluate the energy.

One can also see in Figure~\ref{fig:mbe:4b:meoh} that it's not clear if the MP2 energy
is converged or if it's accumulating numerical noise.
We observe a similar pattern also for the other systems.
Fortunately, the magnitude of the values is small for the other systems, 
close to 0.03~kJ/mol for ammonia and carbon dioxide and even smaller for methane.
We therefore use the average value from data obtained on the last 5~{\AA} as the estimate 
of the MP2 four-body term and set the uncertainty to the absolute value of the average.

Overall, we see that the HF four-body terms contribute between two and five per cent of the total HF energy
for systems with attractive electrostatic interactions (all apart from methane).
The HF four-body terms are positive for all the systems, partially quenching
the three-body terms which are all negative.
Many-body HF contributions beyond fourth-order are likely to be smaller but probably comparable or smaller than
the uncertainties of the three- and four-body contributions.

Even with the large relative uncertainties in the MP2 four-body terms, they represent
only a fraction of per cent of the total MP2 binding energy.
This is not surprising as MP2 accounts only for two electron excitations.
For this reason we expect the MP2 five- and higher-body interactions to be negligible.

\section{Discussion}

\begin{table}[h]
\caption{Hartree-Fock (HF) and MP2 correlation contributions to binding energies of molecular solids in kJ/mol, obtained within
periodic boundary conditions (PBC) and many-body expansion (MBE) performed up to fourth order.
The estimated precision of the PBC values is 0.01~kJ/mol (HF) and 0.1~kJ/mol (MP2).}
\label{tab:dis:hfmp2}
\begin{indented}
\item[]\begin{tabular}{lcccc}
 &\multicolumn{2}{c}{HF} &\multicolumn{2}{c}{MP2}\\
System         &  PBC & MBE & PBC & MBE\\ 
\hline
Methane        & 5.05  & 5.07$\pm0.01$ & $-14.97$ & $-15.01\pm0.02$ \\
Carbon dioxide & $-$3.77 & $-3.78\pm0.07$ & $-25.66$ & $-25.78\pm0.06$\\
Ammonia        & $-$11.09& $-11.01\pm0.16$ & $-23.98$ & $-23.98\pm0.07$\\
Methanol       & $-$18.11& $-18.13\pm0.41$ & $-37.11$ & $-37.15\pm0.13$\\
\end{tabular}
\end{indented}
\end{table}

We collected the final HF and MP2 binding energies obtained using MBE and within PBC in 
Table~\ref{tab:dis:hfmp2}.
One can see that the agreement is excellent for all the studied systems.
The largest deviations, around 0.1~kJ/mol, occur for the HF energy of ammonia and 
for the MP2 energy of carbon dioxide, but the differences are within the estimated uncertainties.

Considering the HF binding energy, the PBC approach is superior to MBE for precise calculations.
The main reason is the simplicity of setting-up the calculations and the fact that
polarization is treated up to infinite order within PBC. 
In contrast, MBE requires the evaluation of a large number of fragments and the non-additive
terms converge slowly.
The reordered summation that we used improves the convergence as it effectively accounts
for interaction with the whole unit cell at once and the electrostatic moments of unit 
cells are typically smaller than the moments of the monomers.
The interaction or non-additive energy is subsequently smaller and the oscillations 
observed for the binding energy as a function of the cut-off distance are reduced as well.

The main contributions to the non-additive terms in HF likely come from polarization
and can be accounted for by polarizable force-fields \cite{wen2011jctc}.
Our preliminary test with a model that accounts only for polarization basically 
confirms this for methanol.
It also shows that our three-body energy is likely converged to within 0.05~kJ/mol, 
much better than our estimated uncertainty of 0.3~kJ/mol.

We found that the main caveat in the HF calculations within PBC is the Coulomb singularity
in the exchange term \cite{gygi1986}.
Fortunately, it can be easily avoided with the Coulomb cut-off scheme \cite{rozzi2006}.
If the cut-off scheme is not used, the energies of solids and molecules need to be extrapolated to infinite volume.
While this is reasonably simple to perform, the extrapolation adds unnecessary uncertainty
to the energies.
Moreover, we used hard PAW data sets which offer high precision for reproducing all-electron
results.
Using standard PAW data sets would introduce an error, especially for systems 
containing oxygen \cite{klimes2016}.

Turning now to MP2 energies, we again find an excellent agreement between the MBE and PBC values
with the differences being only a fraction of percent (see Table~\ref{tab:dis:hfmp2}). 
The largest difference is around 0.5\% for carbon dioxide.
One reason for this is the larger number of electrons in carbon dioxide compared to the other
molecules.
This leads to more demanding calculations, especially within PBC.
For example, we could only use up to 2$\times$2$\times$2 $k$-points for to obtain $E_{\rm cell}^{\rm MP2}$ 
of carbon dioxide while for ammonia with a unit cell of similar size 
a calculation with  3$\times$3$\times$3 $k$-points was possible.

Another issue with the MP2 calculations within PBC is the need to perform several
extrapolations.
Specifically, the extrapolation to infinite volume for the isolated molecule and 
the extrapolation with the basis-set size were problematic.
However, there is one subtle positive point  about the PBC calculations based on plane-waves.
Unlike the gaussian basis sets which are defined for a specific small set of cardinal 
numbers, the size of the plane-wave basis-set can be chosen at will.
For example, we found that the dependence of the binding energy on the basis-set is
noisy for methane and carbon dioxide. 
In such cases, one can compute additional points between the upper and lower limits
to try to improve the quality of the fit.

The MBE calculations require more human time to set-up the computer scripts compared to PBC.
However, once the set-up is automated the small size and embarrassing parallelizability 
of the calculations become benefits.
The MP2 energy converges quickly with the order of MBE and the cutoff distance, even though 
the correlations are considered ``long-range".
Compared to the PBC results, the uncertainties are smaller for MBE-based MP2, apart from methanol.
Due to these reasons, the MBE scheme is preferable for MP2.
One possible issue are the numerical errors which can reach similar magnitude as the data \cite{richard2014}.
This could be avoided by using an accurate model for the interactions
and evaluating it with sufficient numerical precision.

The use of explicit correlation (F12) for MP2 was one of the main benefits of the MBE calculations.
This avoids the need to perform extrapolations which is a substantial simplification,
especially for the user.
Interestingly, the F12 was the most important for short-range dimers, canonical MP2
was sufficient for long-distance dimers and non-additive terms.
One can expect that the F12 scheme will offer similar simplifications of the computational
workflow also within PBC \cite{grueneis2013}.
Another simplification would come from reduction of the memory or compute time
requirements of the periodic MP2 calculations.
This could be accomplished by, {\it e.g.}, using natural orbitals\cite{grueneis2011}
or by using different MP2 algorithms \cite{ochi2015,schafer2017}.

We now compare our results to other MP2 data in the literature.
We find a very good agreement with the binding energies of carbon dioxide obtained
using periodic LMP2 \cite{maschio2010} and MP2 within the hybrid many-body interaction (HMBI) scheme
\cite{wen2011jctc}, see Table~\ref{tab:compar}. 
For ammonia, the agreement is again very satisfactory with the LMP2 value but we find
a larger discrepancy with the HMBI result.
We observe a difference of several kJ/mol for methanol \cite{cervinka2016}, 
but note that our values do not contain monomer relaxation energies which would reduce the difference.
Further differences, of several tenths of kJ/mol, can occur due to the use of different structures
for all the systems.

Finally, we compare our MP2 values to data obtained with other correlated methods. 
Specifically to binding energies obtained using RPA with $GW$ singles corrections \cite{klimes2016},  
CCSD(T) \cite{cervinka2016,wen2011jctc,cervinka2018}, and QMC \cite{zen2018}, see Table~\ref{tab:compar}.
MP2 is in a very good agreement with the reference values for carbon dioxide, 
the accuracy of RPA+GWSE is lower in this case.
The agreement between MP2 and CCSD(T) is worse for ammonia
and methanol. 
However, the CCSD(T) value for ammonia was obtained with the HBMI scheme and we see that
the MP2 binding energy obtained using this approach differs from our
value by around 4~kJ/mol.
Therefore, the MP2 error for ammonia is likely smaller and, in fact, our MP2 binding energy
differs by around 2~kJ/mol from the QMC value of Zen et al.~\cite{zen2018}.
We expect that the disagreement between our MP2 value and the CCSD(T) data
for methanol can be due to similar reasons.

\begin{table}[h]
\caption{The total MP2 binding energies obtained using PBC and MBE methods.
We also list data obtained previously for ammonia and carbon dioxide using periodic 
local MP2 \cite{maschio2010} and by hybrid many-body interaction (HMBI) scheme at MP2 level \cite{wen2011jctc}.
The HMBI value for methanol is from Ref.~\cite{cervinka2017methanol}.
The results for RPA+GWSE calculations are from Ref.~\cite{klimes2016}.
The CCSD(T) values are from Ref.~\cite{cervinka2016} for methane, 
from Ref.~\cite{wen2011jctc} for carbon dioxide and ammonia and 
from Ref.~\cite{cervinka2018} for methanol.
The QMC values are from Ref.~\cite{zen2018}.
}
\label{tab:compar}
\begin{indented}
\item[]\begin{tabular}{lccccccc}
System         &  PBC    & MBE              & LMP2    & HMBI    &RPA+GWSE & CCSD(T) & QMC \\ 
\hline
Methane        & $-9.92$ & $-9.94\pm0.03$   & --      & --      & --     & $-10.97$ &  -- \\
Carbon dioxide & $-29.43$ & $-29.56\pm0.13$ & $-29.8$ & $-29.1$ &$-27.3$ & $-29.5$  & $-28.2\pm1.3$\\
Ammonia        & $-35.07$ & $-34.99\pm0.23$ & $-35.5$ & $-39.3$ &$-37.6$ & $-40.2$  & $-37.1\pm0.4$\\
Methanol       & $-55.22$ & $-55.28\pm0.54$ & --      & $-52.93$& --     & $-51.66$ & --  \\
\end{tabular}
\end{indented}
\end{table}

\section{Conclusions}

We presented a detailed evaluation of HF and MP2 binding energies of four molecular 
solids within periodic boundary conditions and using many-body expansion.
We tried to obtain the contributions with high precision to understand the significant sources of errors.
The values obtained with the two approaches agree well even though 
some of them have higher uncertainties.

Within PBC, the use of the Coulomb cut-off technique was crucial to speed
up the convergence of both HF and MP2 energies with respect to volume 
and $k$-points.
Otherwise the computational and memory requirements are currently limiting, especially for MP2.

The convergence of MBE with real-space cut-off deteriorated with increasing
importance of long-range electrostatic interactions, as expected.
We reduced this issue by reordering the summation, adding some terms that
would otherwise contribute at higher cut-offs.
Numerical noise appeared already for two-body terms, especially for large 
basis sets. 
This should make one cautious when fitting force-fields on energies
obtained for large distances.

Finally, it's simpler to evaluate the HF energy within PBC 
while the situation is less clear cut for MP2.
Currently, the combination of lower level energies obtained for the whole system with MBE corrections 
seems to be the most efficient approach for obtaining binding energies for post-HF methods. 
This is supported by examples in the literature using it both for solids and finite clusters
\cite{tuma2010,muller2013,cervinka2018,modrzejewski2021}.

\section*{Data availability statement}

The data that support the findings of this study are openly available at the following DOI:

10.5281/zenodo.5248078

\section*{Acknowledgements}
This work was supported by the European Union's Horizon 2020 research and innovation programme via the ERC grant APES (No 759721), 
by a PRIMUS project of the Charles University, and by Student faculty grant of the Faculty of Mathematics and Physics
of the Charles University.
We are grateful for computational resources provided by
the IT4Innovations National Supercomputing Center (LM2015070),
CESNET (LM2015042), CERIT Scientific Cloud (LM2015085), and
e-INFRA~CZ (ID: 90140), supported by the Ministry of Education, Youth and Sports
of the Czech Republic.

\section*{References\label{bibby}}
\bibliographystyle{iopart-num}
\bibliography{TS}

\providecommand{\newblock}{}
\begin{thebibliography}{10}
\expandafter\ifx\csname url\endcsname\relax
  \def\url#1{{\tt #1}}\fi
\expandafter\ifx\csname urlprefix\endcsname\relax\def\urlprefix{URL }\fi
\providecommand{\eprint}[2][]{\url{#2}}

\bibitem{price2009}
Price S~L 2009 {\em Acc. Chem. Res.\/} {\bf 42} {117}

\bibitem{reilly2016}
Reilly A~M, Cooper R~I, Adjiman C~S, Bhattacharya S, Boese A~D, Brandenburg
  J~G, Bygrave P~J, Bylsma R, Campbell J~E, Car R, Case D~H, Chadha R, Cole
  J~C, Cosburn K, Cuppen H~M, Curtis F, Day G~M, DiStasio~Jr R~A, Dzyabchenko
  A, van Eijck B~P, Elking D~M, van~den Ende J~A, Facelli J~C, Ferraro M~B,
  Fusti-Molnar L, Gatsiou C~A, Gee T~S, de~Gelder R, Ghiringhelli L~M, Goto H,
  Grimme S, Guo R, Hofmann D~W~M, Hoja J, Hylton R~K, Iuzzolino L, Jankiewicz
  W, de~Jong D~T, Kendrick J, de~Klerk N~J~J, Ko H~Y, Kuleshova L~N, Li X,
  Lohani S, Leusen F~J~J, Lund A~M, Lv J, Ma Y, Marom N, Masunov A~E, McCabe P,
  McMahon D~P, Meekes H, Metz M~P, Misquitta A~J, Mohamed S, Monserrat B, Needs
  R~J, Neumann M~A, Nyman J, Obata S, Oberhofer H, Oganov A~R, Orendt A~M,
  Pagola G~I, Pantelides C~C, Pickard C~J, Podeszwa R, Price L~S, Price S~L,
  Pulido A, Read M~G, Reuter K, Schneider E, Schober C, Shields G~P, Singh P,
  Sugden I~J, Szalewicz K, Taylor C~R, Tkatchenko A, Tuckerman M~E, Vacarro F,
  Vasileiadis M, Vazquez-Mayagoitia A, Vogt L, Wang Y, Watson R~E, de~Wijs G~A,
  Yang J, Zhu Q and Groom C~R 2016 {\em Acta Crystallographica Section B\/}
  {\bf 72} 439--459

\bibitem{santra2013}
Santra B, Klime\v{s} J, Tkatchenko A, Alf\`{e} D, Slater B, Michaelides A, Car
  R and Scheffler M 2013 {\em J. Chem. Phys.\/} {\bf 139} 154702

\bibitem{reilly2013}
Reilly A~M and Tkatchenko A 2013 {\em J. Chem. Phys. Lett.\/} {\bf 4} 1028

\bibitem{klimes2016}
Klime\v{s} J 2016 {\em J. Chem. Phys.\/} {\bf 145} {094506}

\bibitem{delben2012jctc}
Ben M~D, Hutter J and VandeVondele J 2012 {\em J. Chem. Theo. Comput.\/} {\bf
  8} 4177

\bibitem{li2010rpa}
Li Y, Lu D, Nguyen H~V and Galli G 2010 {\em J Phys. Chem. A\/} {\bf 114} 1944

\bibitem{macher2014}
Macher M, Klime\v{s} J, Franchini C and Kresse G 2014 {\em J. Chem. Phys.\/}
  {\bf 140} {084502}

\bibitem{klimes2015}
Klime\v{s} J, Kaltak M, Maggio E and Kresse G 2015 {\em J. Chem. Phys.\/} {\bf
  143} {102816}

\bibitem{booth2013}
Booth H~G, Gr\"uneis A, Kresse G and Alavi A 2013 {\em Nature\/} {\bf 493}
  {365}

\bibitem{mcclain2017}
McClain J, Sun Q, Chan G~K~L and Berkelbach T~C 2017 {\em J. Chem. Theory
  Comput.\/} {\bf 13} 1209--1218

\bibitem{liao2021}
Liao K, Shen T, Li X~Z, Alavi A and Gr\"uneis A 2021 {\em Phys. Rev. B\/} {\bf
  103}(5) 054111

\bibitem{zen2018}
Zen A, Brandenburg J~G, Klime{\v s} J, Tkatchenko A, Alf{\`e} D and Michaelides
  A 2018 {\em Proc. Nat. Acad. Sci. U. S. A.\/} {\bf 115} 1724--1729

\bibitem{gillan2013}
Alf\`e D, Bart\'ok A~P, Cs\'anyi G and Gillan M~J 2013 {\em J. Chem. Phys.\/}
  {\bf 138} 221102

\bibitem{yang2014bz}
Yang J, Hu W, Usvyat D, Matthews D, Sch\"{u}tz M and Chan G~K 2014 {\em
  Science\/} {\bf 345} 640

\bibitem{hartman2016}
Hartman J~D, Kudla R~A, Day G~M, Mueller L~J and Beran G~J~O 2016 {\em Phys.
  Chem. Chem. Phys.\/} {\bf 18}(31) 21686--21709

\bibitem{richard2014acc}
Richard R~M, Lao K~U and Herbert J~M 2014 {\em Acc. Chem. Res.\/} {\bf 47}
  2828--2836

\bibitem{dahlke2007}
Dahlke E~E and Truhlar D~G 2007 {\em J. Chem. Theory Comput.\/} {\bf 3} 46--53

\bibitem{bygrave2012}
Bygrave P~J, Allan N~L and Manby F~R 2012 {\em J. Chem. Phys.\/} {\bf 137}
  164102

\bibitem{wen2011jctc}
Wen S and Beran G~J~O 2011 {\em J. Chem. Theory Comput.\/} {\bf 7} 3733--3742

\bibitem{cervinka2018}
\v{C}ervinka C and Beran G~J~O 2018 {\em Chem. Sci.\/} {\bf 9} 4622–4629

\bibitem{bludsky2008prb}
Bludsk\'y O, Rube\v{s} M and Sold\'an P 2008 {\em Phys. Rev. B\/} {\bf 77}
  {092103}

\bibitem{taylor2012}
Taylor C~R, Bygrave P~J, Hart J~N, Allan N~L and Manby F~R 2012 {\em Phys.
  Chem. Chem. Phys.\/} {\bf 14} 7739

\bibitem{muller2013}
M\"{u}ller C and Usvyat D 2013 {\em J. Chem. Theory Comput.\/} {\bf 9} 5590

\bibitem{beran2016}
Beran G~J~O 2016 {\em Chem. Rev.\/} {\bf 116} 5567

\bibitem{grazulis2009}
Grazulis S, Chateigner D, Downs R~T, Yokochi A~F~T, Quiros M, Lutterotti L,
  Manakova E, Butkus J, Moeck P and Le~Bail A 2009 {\em J. Appl.
  Crystallogr.\/} {\bf 42} 726

\bibitem{cod}
\url{www.crystallography.net/cod/}

\bibitem{boese1997}
Boese R, Niederpr\"{u}m N, Bl\"{a}ser D, Maulitz A, Antipin M~Y and Mallinson
  P~R 1997 {\em J. Phys. Chem. B\/} {\bf 101} 5794

\bibitem{hewat1979}
Hewat A~W and Riekel C 1979 {\em Acta Cryst. A\/} {\bf 35} 569--571

\bibitem{crc}
Haynes W 2016 {\em CRC Handbook of Chemistry and Physics, 97th ed.\/} (CRC
  Press, Boca Raton, FL)

\bibitem{simon1980}
Simon A and Peters K 1980 {\em Acta Crystallogr. B\/} {\bf 36} 2750

\bibitem{torrie1989}
Torrie B, Weng S~X and Powell B 1989 {\em Mol. Phys.\/} {\bf 67} 575--581

\bibitem{kresse1993}
Kresse G and Hafner J 1993 {\em Phys. Rev. B\/} {\bf 47} 558

\bibitem{kresse1999}
Kresse G and Joubert J 1999 {\em Phys. Rev. B\/} {\bf 59} 1758

\bibitem{paier2005}
Paier J, Hirschl R, Marsman M and Kresse G 2005 {\em J. Chem. Phys.\/} {\bf
  122} 234102

\bibitem{paier2006}
Paier J, Marsman M, Hummer K, Kresse G, Gerber I~C and \'Angy\'an J~G 2006 {\em
  J. Chem. Phys.\/} {\bf 124} 154709

\bibitem{marsman2009}
Marsman M, Gr\"uneis A, Paier J and Kresse G {2009} {\em J. Chem. Phys.\/} {\bf
  {130}} {184103}

\bibitem{grueneis2010}
Gr\"uneis A, Marsman M and Kresse G 2010 {\em J. Chem. Phys.\/} {\bf 133}
  {074107}

\bibitem{harl2008}
Harl J and Kresse G 2008 {\em Phys. Rev. B\/} {\bf 77} 045136

\bibitem{kaltak2014rpa1}
Kaltak M, Klime\v{s} J and Kresse G 2014 {\em J. Chem. Theory Comput.\/} {\bf
  10} {2498}

\bibitem{kaltak2014rpa2}
Kaltak M, Klime\v{s} J and Kresse G 2014 {\em Phys. Rev. B\/} {\bf 90} {054115}

\bibitem{rozzi2006}
Rozzi C~A, Varsano D, Marini A, Gross E~K~U and Rubio A 2006 {\em Phys. Rev.
  B\/} {\bf 73} 205119

\bibitem{rupp2012}
Rupp M, Tkatchenko A, M\"uller K~R and von Lilienfeld O~A 2012 {\em Phys. Rev.
  Lett.\/} {\bf 108}(5) 058301

\bibitem{borca2019}
Borca C~H, Bakr B~W, Burns L~A and Sherrill C~D 2019 {\em J. Chem. Phys.\/}
  {\bf 151} 144103

\bibitem{werner2020molpro}
Werner H~J, Knowles P~J, Manby F~R, Black J~A, Doll K, Heßelmann A, Kats D,
  Köhn A, Korona T, Kreplin D~A, Ma Q, Miller T~F, Mitrushchenkov A, Peterson
  K~A, Polyak I, Rauhut G and Sibaev M 2020 {\em J. Chem. Phys.\/} {\bf 152}
  144107

\bibitem{kendall1992}
Kendall R~A, Dunning T~H and Harrison R~J 1992 {\em J. Chem. Phys.\/} {\bf 96}
  6796

\bibitem{kutzelnigg1991A}
Kutzelnigg W and Klopper W {1991} {\em {J. Chem. Phys.}\/} {\bf {94}} {1985}

\bibitem{werner2007}
Werner H~J, Adler T~B and Manby F~R 2007 {\em J. Chem. Phys.\/} {\bf 126}
  164102

\bibitem{adler2007simple}
Adler T~B, Knizia G and Werner H~J 2002 {\em J. Chem. Phys.\/} {\bf 116} 3175

\bibitem{knizia2009simplified}
Knizia G and Adler T Band~Werner H~J 2009 {\em J. Chem. Phys.\/} {\bf 130}
  054104

\bibitem{noga2009on}
Noga J and Šimunek J 2009 {\em Chem. Phys.\/} {\bf 356} 1--6

\bibitem{weigend2002efficient}
Weigend F, Kohn A and Hattig C 2002 {\em J. Chem. Phys.\/} {\bf 127} 221106

\bibitem{weigend2005balanced}
Weigend F and Ahlrichs R 2005 {\em Phys. Chem. Chem. Phys.\/} {\bf 7}
  3297--3305

\bibitem{yousaf2009optimized}
Yousaf K~E and Peterson K~A 2009 {\em Chem. Phys. Lett.\/} {\bf 476} 303--307

\bibitem{gora2011}
Góra U, Podeszwa R, Cencek W and Szalewicz K 2011 {\em J. Chem. Phys.\/} {\bf
  135} 224102

\bibitem{rezac2015}
\v{R}ez\'{a}\v{c} J, Huang Y, Hobza P and Beran G~J~O 2015 {\em J. Chem. Theory
  Comput.\/} {\bf 11} 3065

\bibitem{modrzejewski2021}
Modrzejewski M, Yourdkhani S, Śmiga S and Klime\v{s} J 2021 {\em J. Chem.
  Theory Comput.\/} {\bf 17} 804--817

\bibitem{richard2014}
Richard R~M, Lao K~U and Herbert J~M 2014 {\em J. Chem. Phys.\/} {\bf 141}
  014108

\bibitem{gygi1986}
Gygi F and Baldereschi A {1986} {\em Phys. Rev. B\/} {\bf {34}} {4405}

\bibitem{gajdos2006}
Gajdo\v{s} M, Hummer K, Kresse G, Furthm\"{u}ller J and Bechstedt F {2006} {\em
  Phys. Rev. B\/} {\bf {73}} {045112}

\bibitem{liao2016}
Liao K and Gr\"{u}neis A 2016 {\em J. Chem. Phys.\/} {\bf 145} 141102

\bibitem{gulans_unp}
Gulans A 2014 {\em J. Chem. Phys.\/} {\bf 141} {164127}

\bibitem{klimes2014NC}
Klime\v{s} J, Kaltak M and Kresse G 2014 {\em Phys. Rev. B\/} {\bf 90} {075125}

\bibitem{beran2010jpcl}
Beran G~J~O and Wen S 2010 {\em J. Phys. Chem. Lett.\/} {\bf 1} 3480--3487

\bibitem{takatani2010}
Takatani T, Hohenstein E~G, Malagoli M, Marshall M~S and Sherrill C~D 2010 {\em
  J. Chem. Phys.\/} {\bf 132} 144104

\bibitem{kennedy2014}
Kennedy M~R, McDonald A~R, DePrince A~E, Marshall M~S, Podeszwa R and Sherrill
  C~D 2014 {\em J. Chem. Phys.\/} {\bf 140} 121104

\bibitem{heindel2020}
Heindel J and Xantheas S 2020 {\em J. Chem. Theory Comput.\/} {\bf 16}
  6843--6855

\bibitem{grueneis2013}
Gr\"uneis A, Shepherd J~J, Alavi A, Tew D~P and Booth G~H 2013 {\em J. Chem.
  Phys.\/} {\bf 139} {084112}

\bibitem{grueneis2011}
Gr\"uneis A, Booth G~H, Marsman M, Spencer J, Alavi A and Kresse G 2011 {\em J.
  Chem. Theory Comput.\/} {\bf 7} {2780}

\bibitem{ochi2015}
Ochi M and Tsuneyuki S 2015 {\em Chem. Phys. Lett.\/} {\bf 621} 177--183 ISSN
  0009-2614

\bibitem{schafer2017}
Sch\"{a}fer T, Ramberger B and Kresse G 2017 {\em J. Chem. Phys.\/} {\bf 146}
  104101

\bibitem{maschio2010}
Maschio L, Usvyat D, Sch\"{u}tz M and Civalleri B {2010} {\em J. Chem. Phys.\/}
  {\bf {132}} {134706}

\bibitem{cervinka2016}
\v{C}ervinka C, Fulem M and Ru\v{z}i\v{c}ka K 2016 {\em J. Chem. Phys.\/} {\bf
  144} 164505

\bibitem{cervinka2017methanol}
Červinka C and Beran G~J~O 2017 {\em Phys. Chem. Chem. Phys.\/} {\bf 19}(44)
  29940--29953

\bibitem{tuma2010}
Tuma C, Kerber T and Sauer J 2010 {\em Angew. Chem. Int. Ed.\/} {\bf 49} 4678

\end{thebibliography}
\end{document}